# Demonstration of the Relationship between Sensitivity and Identifiability for Inverse Uncertainty Quantification

Xu Wu[a,*], Koroush Shirvan[a] and Tomasz Kozlowski[b]

[a]Department of Nuclear Science and Engineering, Massachusetts Institute of Technology, Cambridge, MA, USA

[b]Department of Nuclear, Plasma and Radiological Engineering, University of Illinois at Urbana-Champaign, Urbana, IL, USA

Corresponding author: (+1) 217-979-7432

Emails: xuwu@mit.edu (X. Wu), kshirvan@mit.edu (K. Shirvan), txk@illinois.edu (T. Kozlowski)

**Abstract**

Inverse Uncertainty Quantification (UQ), or Bayesian calibration, is the process to quantify the uncertainties of random input parameters based on experimental data. The introduction of model discrepancy term is significant because "over-fitting" can theoretically be avoided. But it also poses challenges in the practical applications. One of the mostly concerned and unresolved problem is the "lack of identifiability" issue. With the presence of model discrepancy, inverse UQ becomes "non-identifiable" in the sense that it is difficult to precisely distinguish between the parameter uncertainties and model discrepancy when estimating the calibration parameters. Previous research to alleviate the non-identifiability issue focused on using informative priors for the calibration parameters and the model discrepancy, which is usually not a viable solution because one rarely has such accurate and informative prior knowledge. In this work, we show that identifiability is largely related to the sensitivity of the calibration parameters with regards to the chosen responses. We adopted an improved modular Bayesian approach for inverse UQ that does not require priors for the model discrepancy term. The relationship between sensitivity and identifiability was demonstrated with a practical example in nuclear engineering. It was shown that, in order for a certain calibration parameter to be statistically identifiable, it should be significant to at least one of the responses whose data are used for inverse UQ. Good identifiability cannot be achieved for a certain calibration parameter if it is not significant to any of the responses. It is also demonstrated that "fake identifiability" is possible if model responses are not appropriately chosen, or inaccurate but informative priors are specified.

## 1. Introduction

Inverse Uncertainty Quantification (UQ), also referred to as *inverse/backward problem* [1] or *parameter estimation*, is the process to quantify the uncertainties of random input parameters based on experimental data. Inverse UQ is crucial to forward uncertainty propagation, sensitivity and validation studies which all suffer from the "lack of input uncertainty information" issue. It seeks uncertainties in calibration parameters that are most consistent with the physical observations, the computer model and any prior beliefs gained through previous experiments or expert judgments. Inverse UQ by definition is very similar to *Bayesian/statistical calibration* [2][3], which is also known as *Calibration under Uncertainty (CUU)* [4]. In this work we treat both processes as the same and will use them interchangeably.

Most inverse UQ related work of nowadays follow the seminal work of Kennedy and O'Hagan [2], hereafter referred to as the "KOH" approach. The KOH approach is comprehensive in that it accounts for many sources of uncertainties, especially model discrepancy. *Model discrepancy*, also referred to as *model inadequacy*, *model uncertainty*, *model bias*, *model error*, *model form error* or *structural uncertainty*, is due to insufficient or inaccurate underlying physics, numerical approximation errors, and/or other inaccuracies that would exist even if all the parameters in the computer model were known [2][5][6]. It is important to consider model discrepancy as otherwise we would have an unrealistic level of confidence in the computer simulations [5].

Inverse UQ is known to be conceptually and mathematically much more difficult than forward UQ [1], mainly because of the ill-posedness [7] of inverse problems. The introduction of model discrepancy term by the KOH approach is significant because "over-fitting" (biased parameter estimations, which means that the calibration parameters are over-calibrated to a certain set of observation data, causing large prediction errors when the computer model is used for other experimental conditions) during inverse UQ can theoretically be avoided [5]. But it also poses challenges in the practical applications. One of the mostly concerned and unresolved problem is the "lack of identifiability" issue [5][6][8]. Identifiability refers to whether the true values of calibration parameters can theoretically be inferred based on given data. Without the model discrepancy term, inverse UQ is a straightforward and statistically "identifiable" process. However, with the presence of model discrepancy, inverse UQ becomes "non-identifiable" in the sense that it is difficult to precisely estimate the calibration parameters and to distinguish between the effects of parameter uncertainty and model discrepancy.

The root of non-identifiability is the confounding between the calibration parameters and the model discrepancy [6][9][10]. Multiple different combinations of the computer model (with different values for the calibration parameters) and its corresponding model discrepancy might result in equally good agreement with the measurement data and equally high values for the likelihood function during Markov Chain Monte Carlo (MCMC) sampling. For instance, Jiang et al. [11] illustrated a simple case in which three different combinations of calibration parameter and discrepancy function result in equally good agreement with the physical observations.

The degree of identifiability is measured by the posterior standard deviation (in case of single) [6][8] or posterior covariance matrix (in case of multiple) [11][12] of the calibration parameters. A tight posterior distribution indicates that a parameter is identifiable while a widely dispersed one means that a parameter cannot be precisely quantified and hence has poor identifiability. It is crucial to achieve good identifiability during inverse UQ for the following reasons:

1. The primary goal of inverse UQ is to address the "lack of input uncertainty information" issue. Uncertainties of calibration parameters can only be successfully learned with good identifiability.

2. Good identifiability also yields accurate estimation of model discrepancy posterior, which helps determining model deficiencies and provides guidance for future improvement of computer models.

3. Better knowledge of the calibration parameters produces more accurate computer model predictions over a broad domain of application. That is, most calibration parameters have fixed "true" values which are independent of the context of the simulator's application [5]. The posteriors learned from the calibration domain are still applicable for domains where no physical observations are available, e.g. the extrapolated prediction domain.

The most popular and tested method to describe model discrepancy is Gaussian Process (GP) [13][14] following the KOH approach [2][15]. In order to deal with the lack of identifiability, it is recommended in [5][9][10] that one should use proper prior distributions (informative, usually with a specific functional form) for the calibration parameters and the model discrepancy function. However, it is often impossible to assign informative priors for either calibration parameters or model discrepancy because one rarely has significant prior knowledge of them. Firstly, calibration parameters with informative priors would have been treated as known rather than targets for inverse UQ. Secondly, it is inherently paradoxical to find a proper functional form of model discrepancy when one does not know the reality. Finally, it was demonstrated in [6] that there is inherent danger of using informative but inaccurate priors. Such misspecified priors can possibly result in tight posterior distributions for calibration parameters which are "far away" from the "true" solutions. In this case, one will mistakenly believe that good identifiability has been achieved while the solutions are actually wrong.

Ling and colleagues [16] investigated five different prior formulations of model discrepancy function for Bayesian calibration: (1) constant, (2) Gaussian random variable with fixed mean and variance, (3) Gaussian random variable with input-dependent mean and variance, (4) Gaussian random process with stationary covariance function, and (5) Gaussian random process with non-stationary covariance function. Five posterior distributions of calibration parameters and model discrepancy were obtained based on the five priors. Next a reliability-based validation metric was used to assess the model predictions using each of the posteriors. The resulting quantitative validation metrics then served as weights to combine the five posterior distributions into a single distribution. Such Bayesian model averaging accounts for the uncertainty induced by the lack of knowledge regarding the model discrepancy. However, this method has limitations in practical applications. Firstly, some of the priors may not reflect the true model discrepancy and they may lead to wrong posteriors. The validation step may not be able to rule the corresponding posterior out (by assigning small weights in Bayesian model averaging) especially when validation data is limited.

Secondly, there is no guidance in choosing an appropriate threshold for the reliability-based validation metric. Thirdly, this technique may be useful to reduce the contribution from an improper prior for model discrepancy, but the proposed five priors does not represent all possible functional forms. For instance, model discrepancy may be linear or quadratic functions. Finally, considering multiple functional forms for priors will induce large computational cost when there are multiple responses.

Arendt and colleagues [8] showed that good identifiability can be achieved by including multiple responses that share a mutual dependence on a common set of calibration parameters. The authors extended the single response modular Bayesian approach [6] to multiple responses for calculating posteriors of the calibration parameters and the model discrepancy. With numerical study of a simply supported beam, it was demonstrated that considering multiple responses can improve identifiability by an amount that ranges from minimal to substantial, depending on the characteristics of the specific responses that were combined. However, considering each combination one-by-one may be computationally prohibitive for problems with many responses. Including all of the responses is also not a viable solution due to the following reasons: (1) including additional responses increases the computational cost and may cause numerical instabilities (e.g. conditioning issue), (2) some responses may be redundant in the sense that they contain the same information as others, (3) many responses are easily available from computer simulations but not measurable or prohibitively costly to measure in reality.

More recent work [11][12] addressed the issue of how to select the most appropriate subset of responses to best enhance identifiability. The authors used a preposterior analysis approach that, prior to conducting the physical experiments but after conducting the computer simulations, can predict the relative improvement in identifiability that will result using different subsets of responses. However, the preposterior analysis approach induces large computational cost. Its application in practice is also constrained by cost in physical experiments. For example, in nuclear engineering, legacy data are usually used for inverse UQ since new experiments are normally expensive, which means that we are constrained to only use responses with available data.

The work presented in this paper is inspired by the work presented in [6][8]. The authors showed in [6] that the identifiability strongly depends on the "nature" of the computer model responses as a function of the calibration parameters. It was also pointed in [8] that different combinations of responses result in drastically different identifiability. However, it was not explained how the "nature" of model responses as functions of the calibration parameters affect the identifiability for a generic problem. Better understanding of why certain combinations of multiple responses improve identifiability much more than others is necessary, which can guide the users to systematically choose the responses that result in the largest improvement in identifiability.

In this work, we will show that identifiability is largely related to the sensitivity of the calibration parameters w.r.t. the chosen responses. In order for a certain calibration parameter to be statistically identifiable, it should be significant to at least one of the responses whose observation data are used for inverse UQ. To the best of the authors' knowledge, no previous work has studied the connection between sensitivity and identifiability. Sensitivity Analysis (SA) is the study of how uncertainties in the responses can be attributed to various input parameters [17]. SA provides a ranking of the input parameters by their importance to responses. In this work we use Sobol' indices to represent sensitivity. Sobol' indices [18] measures the percentage of the variances (or uncertainties) in responses that can be apportioned to each one of the input parameters or their combinations, which is a very straightforward measure of sensitivity.

We investigated a practical example in nuclear engineering in which the calibration parameters have distinct sensitivities for the responses. The numerical test is the inverse UQ of nuclear reactor system thermal-hydraulics code TRACE [19] physical model parameters, using experimental data from the OECD/NEA BWR Full-size Fine-Mesh Bundle Tests (BFBT) benchmark [20]. This test consists of five calibration parameters and four responses. Each response has different significant contributors. We adopted an improved modular Bayesian approach developed in a recent work [13][21] for inverse UQ which does not require priors for the model discrepancy GP emulator. By performing inverse UQ with different combinations of the responses and looking at the corresponding posteriors, a connection between sensitivity and identifiability can be established.

This paper is organized in the following way. Section 2 briefly discusses the inverse UQ methodology and the identifiability issue. Section 3 presents a general overview of global sensitivity analysis using Sobol' indices. The numerical test example is defined in Section 4. Sections 5 and 6 present the results. Section 5 concludes the paper.

## 2. Inverse Uncertainty Quantification

### 2.1. Bayesian formulation for inverse UQ

In this section, we will briefly present the Bayesian formulation for inverse UQ problems. Consider a general computer model $\mathbf{y}^M = \mathbf{y}^M(\mathbf{x}, \boldsymbol{\theta})$ where $\mathbf{y}^M$ is the model response, $\mathbf{x}$ is the vector of design variables, and $\boldsymbol{\theta}$ is the vector of calibration parameters (See [13] for detailed discussion on the classification of input parameters). Denote the physically measured response as $\mathbf{y}^E(\mathbf{x})$, the "model updating equation" [2][6] is defined as:

$$\mathbf{y}^E(\mathbf{x}) = \mathbf{y}^M(\mathbf{x}, \boldsymbol{\theta}^*) + \delta(\mathbf{x}) + \boldsymbol{\varepsilon} \tag{1}$$

where $\boldsymbol{\theta}^*$ represents the "true" but unknown values for the calibration parameters, the learning of which is the goal of inverse UQ process. The term $\delta(\mathbf{x})$ is called model discrepancy [2], which is due to incomplete or inaccurate underlying physics, numerical approximation errors, and/or other inaccuracies that would exist even if all the parameters in the computer model were known. Finally $\boldsymbol{\varepsilon} \sim \mathcal{N}(\mathbf{0}, \boldsymbol{\Sigma}_{\text{exp}})$ represents the measurement noises. The model updating equation serves as the starting point of the inverse UQ process. The model discrepancy term was first addressed in the seminal work of Kennedy and O'Hagan [2]. Based on the model updating equation and the Gaussian assumption of the measurement noises, $\boldsymbol{\varepsilon} = \mathbf{y}^E(\mathbf{x}) - \mathbf{y}^M(\mathbf{x}, \boldsymbol{\theta}^*) - \delta(\mathbf{x})$ follows a multi-dimensional Gaussian distribution. The posterior can be written as:

$$p(\boldsymbol{\theta}^* | \mathbf{y}^E, \mathbf{y}^M) \propto p(\boldsymbol{\theta}^*) \cdot \frac{1}{\sqrt{|\boldsymbol{\Sigma}|}} \exp\left[-\frac{1}{2} [\mathbf{y}^E - \mathbf{y}^M - \delta]^T \boldsymbol{\Sigma}^{-1} [\mathbf{y}^E - \mathbf{y}^M - \delta]\right] \tag{2}$$

Note that the likelihood covariance matrix $\boldsymbol{\Sigma}$ has three parts:

$$\boldsymbol{\Sigma} = \boldsymbol{\Sigma}_{\text{exp}} + \boldsymbol{\Sigma}_{\text{bias}} + \boldsymbol{\Sigma}_{\text{code}} \tag{3}$$

The first term $\boldsymbol{\Sigma}_{\text{exp}}$ is the *experimental uncertainty* caused by measurement noise $\boldsymbol{\varepsilon}$. The second term $\boldsymbol{\Sigma}_{\text{bias}}$ represents the *model uncertainty*, as stated earlier, due to incomplete/inaccurate underlying physics and numerical approximation errors. The third term $\boldsymbol{\Sigma}_{\text{code}}$ is called *code uncertainty*, or *interpolation uncertainty*, because we do not know the computer code outputs at every input setting, especially when the code is computationally prohibitive. In this case, one might choose to use some kind of metamodels (e.g. GP). Code uncertainty should only be considered when metamodels are used. For a complete and detailed discussion of the inverse UQ formulation, see [13]. The posterior function can be explored by MCMC sampling to obtain the posterior distributions.

### 2.2. Treatment of the model discrepancy term

The introduction of model discrepancy term by the KOH approach is comprehensive in the sense that many different sources of uncertainties are considered in Bayesian calibration, i.e. uncertainties from parameter, model, experiment and code. It is also advantageous to consider model discrepancy because "over-fitting" (biased parameter estimates) can theoretically be avoided. However, it also poses challenges in the practical engineering applications. As shown by Brynjarsdottir and O'Hagan [5], the calibration results may not be satisfying if the prior assumption of model discrepancy does not capture the effect of missing physics in the model. However, the missing/insufficient physics is not known or not quantifiable because otherwise model developers could have incorporated them in the computer codes. Therefore, researchers have been trying to describe the model discrepancy term mathematically. One of the most popular choices is using GP [14]. Treatment of model discrepancy with GP can be further classified as two different techniques, full Bayesian [22][23] and modular Bayesian [6][8][10]. As this paper is not intended to review on these approaches, in the following we will only briefly comment on full/modular Bayesian. Interested readers can refer to [13] for a detailed review and comparison.

In brief, both full and modular Bayesian approaches use a GP emulator to replace the full model (original computer code) during MCMC sampling, and a second GP model to represent the model discrepancy. Both GP models have unknown hyperparameters, $\boldsymbol{\Psi}^M = \{\boldsymbol{\beta}^M, \sigma_M^2, \boldsymbol{\omega}^M, \mathbf{p}^M\}$ for computer model and $\boldsymbol{\Psi}^\delta = \{\boldsymbol{\beta}^\delta, \sigma_\delta^2, \boldsymbol{\omega}^\delta, \mathbf{p}^\delta\}$ for model discrepancy, where $\boldsymbol{\beta}$ are the basis functions, $\sigma^2$ are the process variances, $\boldsymbol{\omega}$ are the characteristic length-scales and $\mathbf{p}$ are the roughness parameters [13]. Full and modular Bayesian approaches differ in their treatment of the unknown

GP hyperparameters. In full Bayesian, both $\mathbf{\Psi}^M$ and $\mathbf{\Psi}^\delta$ are treated in a similar way as the calibration parameters $\boldsymbol{\theta}$. They are assigned priors which also enter the likelihood function. Eventually posteriors of $\mathbf{\Psi}^M$, $\mathbf{\Psi}^\delta$ and $\boldsymbol{\theta}$ are solved for all together. Then $\mathbf{\Psi}^M$ and $\mathbf{\Psi}^\delta$ need to be integrated out from the joint posterior to get marginal distributions of $\boldsymbol{\theta}$. However, in modular Bayesian, the estimation of $\boldsymbol{\theta}$, $\mathbf{\Psi}^M$ and $\mathbf{\Psi}^\delta$ are all separated. Modular Bayesian uses plausible estimates of $\{\mathbf{\Psi}^M, \mathbf{\Psi}^\delta\}$ evaluated by methods like Maximum Likelihood Estimation (MLE) and treat them as if they were the true values of $\{\mathbf{\Psi}^M, \mathbf{\Psi}^\delta\}$.

Both full and modular Bayesian approaches suffer greatly from the "lack of identifiability" issue. In the introduction part, discussions on identifiability have been presented concerning its origin, importance, previous treatments and limitations. The most investigated method to resolve the "lack of identifiability" issue is to use appropriate priors for model discrepancy. Note that the "prior for model discrepancy" essentially means the "priors for model discrepancy GP hyperparameters $\mathbf{\Psi}^\delta$". Previously, researchers used distributions like normal, gamma, beta and inverse gamma [9][22][23][24][25], but there were no explanations about how the distribution parameters were chosen.

In this work, we adopt an improved modular Bayesian approach developed in a recent work [13][21] that does not require priors for the model discrepancy term. The improved modular Bayesian approach also uses two GP models to represent full model and model discrepancy respectively. The unknowns $\mathbf{\Psi}^M$ are estimated using training samples from carefully designed full model runs. The primary distinction of the proposed method with classic modular Bayesian approach [6][8][10] is the way of solving for $\mathbf{\Psi}^\delta$. To solve for $\mathbf{\Psi}^\delta$ with MLE, we need training data whose input is $\mathbf{x}$ and output is the "observations" of model discrepancy. However, such direct observations of $\delta(\mathbf{x})$ are never available because the reality is unknown. Therefore, we need substitutes of such "observation data" for model discrepancy. The improved modular Bayesian approach [13] is motivated by finding such "observation data".

This approach first uses a sequential Test Source Allocation (TSA) algorithm [21] to separate given data for *validation* and *calibration (inverse UQ)*. Then the computer model is executed in the *validation domain* using the nominal values or prior means of the calibration parameters. The differences between the resulting model outputs and physical observations serve as "observation data" for model discrepancy, based on which a GP model is trained while the *observation noise* serves as "nugget" term. In the fitting process, $\mathbf{\Psi}^\delta = \{\boldsymbol{\beta}^\delta, \sigma_\delta^2, \boldsymbol{\omega}^\delta, \mathbf{p}^\delta\}$ are estimated using MLE and requires no prior distributions. Evaluating this GP model at design variables $\mathbf{x}$ in the *calibration domain* provides the model discrepancy estimations, which will enter the likelihood function during MCMC sampling.

**2.3. Dealing with the identifiability issue**

The improved modular Bayesian approach [13][21] does not require priors for the model discrepancy. But this does not mean that it is capable of bypassing the identifiability issue. In this work, we will show that identifiability is largely related to the sensitivity of the calibration parameters w.r.t. the chosen responses. The connection between sensitivity and identifiability can be explained in an intuitive way. Suppose there are two calibration parameters and one responses. Parameter A is significant in the sense that its uncertainty causes 99% of the variation in the response, while parameter B can only accounts for the remaining 1%. During the random walk in MCMC sampling, most values in the prior range of parameter B produce equally well-fitting of the model response with observations. Therefore, unlike parameter A, the posterior samples of parameter B will not be concentrated in a certain region, resulting in large posterior standard deviation (STD) and non-identifiability.

In order for a certain calibration parameter to be statistically identifiable, it should be significant to at least one of the responses whose measurement data are used for inverse UQ. If such a condition cannot be satisfied, improving the model discrepancy prior in full Bayesian or classical modular Bayesian approaches will not alleviate the non-identifiability problem. To justify this claim, we applied the improved modular Bayesian approach to a practical problem in nuclear engineering, which is the inverse UQ of TRACE [19] physical model parameters using BFBT benchmark steady-state void fraction data [20]. This test consists of five calibration parameters and four responses with different significant contributors. By performing inverse UQ with different combinations of the responses and looking at the corresponding posteriors, a connection between sensitivity and identifiability can be established. We will also detect the strength of identifiability by the posterior STDs. Note that posterior STD may be an oversimplified indication of the spread of a distribution, especially for non-normal cases. However, it is believed to be the most relevant single measure of identifiability and used in many previous work [6][8][11][12][16].

## 3. Sobol' indices for global sensitivity analysis

In this section, we briefly introduce global sensitivity analysis (SA) using Sobol' indices. SA is the process to determine how uncertain input parameters contribute to the variation in the outputs. Methods for SA can be generally categorized as local or global [17][26]. Local SA methods mainly focus on the variations of the model outputs using derivative-based methods around nominal values of the inputs, whereas global methods deal with the uncertainties of the outputs due to input variations over the whole domain. The variance-based methods for global SA mainly use ANOVA (ANalysis Of VAriance) decomposition which represents the variance of a certain output as a sum of contributions of each one of inputs or their combinations. As a straightforward method to identify significant input parameters, Sobol' method [18][27] is a popular measure and it will be used in the current study.

Following the notation used in [28], define a general computer model as $Y = f(X_1, X_2, \ldots, X_d)$ where $\boldsymbol{X} = \{X_1, X_2, \ldots, X_d\}$ is the vector of all input parameters including both design variables and calibration parameters. Without loss of generality, the model output $Y$ is assumed to be a scalar. Define the variance of $Y$ with respect to a fixed input $X_i$ as $\{\text{Var}_{(X_i)}\{\mathbb{E}_{(\boldsymbol{X}_{\sim i})}(Y|X_i)\}, i = 1,2,\ldots,d\}$, where $\boldsymbol{X}_{\sim i}$ represents all input parameters but $X_i$. The inner expectation operator means that the mean of $Y$ is taken over all possible values of $\boldsymbol{X}_{\sim i}$ given a fixed $X_i$, while the outer variance is taken over all possible values of $X_i$. Define the total variance of $Y$ as $\text{Var}(Y)$, the first order sensitivity coefficient, or Sobol' index is defined as:

$$S_i = \frac{\text{Var}_{(X_i)}\{\mathbb{E}_{(\boldsymbol{X}_{\sim i})}(Y|X_i)\}}{\text{Var}(Y)} \quad (4)$$

The first order Sobol' index $S_i$ is also called the *main effect* and it quantifies the variability in Y that is caused by uncertainty in $X_i$ alone. Furthermore, according to the law of total variation [29]:

$$\mathbb{E}_{(X_i)}\{\text{Var}_{(\boldsymbol{X}_{\sim i})}(Y|X_i)\} = \text{Var}(Y) - \text{Var}_{(X_i)}\{\mathbb{E}_{(\boldsymbol{X}_{\sim i})}(Y|X_i)\} \quad (5)$$

It follows that $\text{Var}_{(X_i)}\{\mathbb{E}_{(\boldsymbol{X}_{\sim i})}(Y|X_i)\}$ must be between 0 and $\text{Var}(Y)$, indicating that $0 \leq S_i \leq 1$. If we flip $\boldsymbol{X}_{\sim i}$ and $X_i$ in Equation (5), we get:

$$\mathbb{E}_{(\boldsymbol{X}_{\sim i})}\{\text{Var}_{(X_i)}(Y|\boldsymbol{X}_{\sim i})\} = \text{Var}(Y) - \text{Var}_{(\boldsymbol{X}_{\sim i})}\{\mathbb{E}_{(X_i)}(Y|\boldsymbol{X}_{\sim i})\} \quad (6)$$

Another kind of sensitivity coefficient naturally forms as:

$$T_i = \frac{\mathbb{E}_{(\boldsymbol{X}_{\sim i})}\{\text{Var}_{(X_i)}(Y|\boldsymbol{X}_{\sim i})\}}{\text{Var}(Y)} = 1 - \frac{\text{Var}_{(\boldsymbol{X}_{\sim i})}\{\mathbb{E}_{(X_i)}(Y|\boldsymbol{X}_{\sim i})\}}{\text{Var}(Y)} \quad (7)$$

Since $\text{Var}_{(\boldsymbol{X}_{\sim i})}\{\mathbb{E}_{(X_i)}(Y|\boldsymbol{X}_{\sim i})\}/\text{Var}(Y)$ on the right hand side of Equation (7) can be treated as the first order sensitivity indices of $\boldsymbol{X}_{\sim i}$, the expression $1 - \text{Var}_{(\boldsymbol{X}_{\sim i})}\{\mathbb{E}_{(X_i)}(Y|\boldsymbol{X}_{\sim i})\}/\text{Var}(Y)$ is the portion of total variance that is caused by all input combinations that include $X_i$. Therefore, $T_i$ measures the *total effect* and is called total sensitivity coefficient. Since $T_i$ includes first order effects of $X_i$ and higher order effects of $X_i$ by interaction with other inputs, it is always larger than (when interactions are non-negligible) or equal to (when interactions are negligible) the main effect $S_i$. The formulation of main and total effects can also be derived from Sobol' decomposition, also called variance decomposition [27][28]. For independent input parameters, the Sobol' indices satisfy the following relation:

$$\sum_{1 \leq i \leq d} S_i + \sum_{1 \leq i < j \leq d} S_{ij} + \sum_{1 \leq i < j < l \leq d} S_{ijl} + \cdots S_{1,2,3,\ldots,d} = 1 \quad (8)$$

There are $2^d$ sensitivity indices in total. Indices with multiple subscripts (e.g. $S_{ij}$) are called *interaction terms*. The total Sobol' index $T_i$ for a given input $X_i$ is the sum of all terms in Equation (8) that contain the subscript $(i)$. For example, when $d = 3$, the total effect of $X_1$ includes four parts $T_1 = S_1 + S_{12} + S_{13} + S_{123}$. As the number of indices grows exponentially with the dimension $d$, it is impractical to compute all the sensitivity indices. Furthermore, it is common for the higher order (larger than two) interaction effects to be negligible [17]. The main and total effects are usually sufficient to identify the significant input parameters, while the second-order interaction effects are occasionally considered. Given the main and total effects, there are two straightforward ways to identify the existence of high-order interactions: (1) if the total effect is sufficiently larger than the main effect for a certain input, interaction of this input with others exists whose magnitude depends on the difference between main/total effects, (2) alternatively, one can look at the sum of main or total effects for all the inputs. If main effects of all the inputs sum to a value less than but close to 1.0, or if the sum of total effects results in a value greater than but close to 1.0, it can be concluded that the interactions are negligible.

Sobol' indices have gained wide interest and several methods have been developed to calculate them [17]. Monte Carlo or Quasi Monte Carlo use brute-force sampling methods [18][27][28][30] but they are hardly applicable for computationally prohibitive models, because hundreds of thousands of computer model runs are usually needed. In case of time-consuming computer models, metamodel-based (e.g. GP [31]) sampling can be used. Other popular approaches include Polynomial Chaos Expansion (PCE) [26][32]. PCE is a method that expands the model outputs with respect to orthogonal polynomials in the uncertain inputs. Based on the orthogonality nature of the polynomials used to construct PCE, the output variance caused by each input and their interactions with others are very straightforward to calculate. In the present work, the DAKOTA code [33] is used to calculate Sobol' indices using PCE-based approach. In case of correlated input parameters, methods developed in [34][35][36][37] can be used.

## 4. Problem Definition: Inverse UQ of TRACE Physical Model Parameters with BFBT Benchmark Data

TRACE [19] is a best-estimate system thermal-hydraulics code that has been widely used in nuclear reactor design and safety analysis. Significant uncertainties exist in the physical model parameters of TRACE closure laws, which are used to describe the transfer terms in the mass, momentum and energy balance equations. In previous uncertainty, sensitivity and validation studies, uncertainties in these parameters are usually ignored or defined using "expert opinion" or "user self-assessment". In the present work, uncertainties in TRACE physical model parameters will be inversely quantified based on steady-state void fraction data from BFBT benchmark.

Cross-sectional averaged void fractions were measured at four different axial locations in BFBT assembly 4, hereafter referred to as VoidF1, VoidF2, VoidF3 and VoidF4 respectively from lower to upper positions. Detailed description of TRACE and BFBT benchmark can be found in [19][20]. In this paper, we will not provide such details to avoid detour in the flow of the narrative. In brief, the responses are VoidF1, VoidF2, VoidF3 and VoidF4. The design variables **x** consists of four parameters: pressure, coolant mass flow rate, power and coolant inlet temperature, whose test ranges are shown in Table 1. These design variables are used to define the inverse UQ domain and validation domain in the improved modular Bayesian approach [13][21].

Table 1. Design variables and their ranges in BFBT benchmark.

| Design variables **x** | Units | Lower test range | Upper test range |
|---|---|---|---|
| Pressure | MPa | 0.9730 | 8.7050 |
| Inlet mass flow rate | kg/s | 2.8000 | 19.3583 |
| Power | MW | 0.2200 | 7.3300 |
| Coolant inlet temperature | K | 440.4929 | 564.5076 |

The calibration parameters **θ** consists of five physical model parameters: P1008, P1012, P1022, P1028 and P1029, as described in Table 2. All of the five calibration parameters are multiplicative factors with nominal values of 1.0. The priors are chosen as uniform distributions over the range of (0, 5) for all the parameters. Such ranges are wide to reflect the ignorance before observing any data. Posterior ranges after inverse UQ are expected to be much narrower than prior ranges, indicating that the knowledge of these parameters has been improved given physical observations.

Table 2. Selected TRACE physical model parameters.

| Calibration parameters **θ** (multiplication factors) | Representations | Uniform ranges | Nominals |
|---|---|---|---|
| Single phase liquid to wall HTC | P1008 | (0.0, 5.0) | 1.0 |
| Subcooled boiling HTC | P1012 | (0.0, 5.0) | 1.0 |
| Wall drag coefficient | P1022 | (0.0, 5.0) | 1.0 |
| Interfacial drag (bubbly/slug Rod Bundle - Bestion) coefficient | P1028 | (0.0, 5.0) | 1.0 |
| Interfacial drag (bubbly/slug Vessel) coefficient | P1029 | (0.0, 5.0) | 1.0 |

# 5. Results for the First Inverse UQ Process

## 5.1. Sensitivity analysis

TRACE includes user access to 36 physical model parameters [19][21]. Those reported in Table 2 are selected after dimensional reduction using local/global SA. The removed 31 physical model parameters are either inactive (cause no variation in the responses at all) or have negligible importance (with main effect Sobol' indices less than 1.0E-06 for any one of the responses) for BFBT benchmark. See [38] for the detailed dimensional reduction process. Table 3 shows the Sobol' indices of each parameter for each response which are calculated using prior distributions in Table 2 and Figure 1 visualizes the results. The observed sensitivity can be briefly described as below:

1) The significance of P1008 decreases to almost zero at higher elevations. This is because single-phase liquid exists only in the lower elevations of the bundle.
2) Similarly, P1012 is more important at lower elevations because that is where subcooled boiling occurs.
3) P1022 increases at higher elevations, while P1028 dominates at intermediate elevations.
4) P1029 is only important for VoidF4.

Table 3: Sobol' indices for selected TRACE physical model parameters, calculated using prior distributions.

| Output | Main effect Sobol' indices | | | | Total effect Sobol' indices | | | |
|---|---|---|---|---|---|---|---|---|
| | VoidF1 | VoidF2 | VoidF3 | VoidF4 | VoidF1 | VoidF2 | VoidF3 | VoidF4 |
| P1008 | 0.0903 | 0.0507 | 0.0093 | 0.0008 | 0.2989 | 0.1107 | 0.0176 | 0.0009 |
| P1012 | 0.6875 | 0.2696 | 0.0320 | 0.0012 | 0.9064 | 0.3307 | 0.0404 | 0.0013 |
| P1022 | 0.0022 | 0.1442 | 0.1896 | 0.6285 | 0.0119 | 0.1490 | 0.2022 | 0.6551 |
| P1028 | 0.0006 | 0.4705 | 0.7483 | 0.0299 | 0.0031 | 0.4747 | 0.7607 | 0.0388 |
| P1029 | 0.0000 | 0.0000 | 0.0000 | 0.3112 | 0.0000 | 0.0000 | 0.0000 | 0.3338 |
| Sum | 0.7806 | 0.9350 | 0.9792 | 0.9715 | 1.2203 | 1.0650 | 1.0208 | 1.0298 |

The fact that a certain input has close main and total effects Sobol' indices means that this input has no interaction with others (e.g. P1022, P1028 and P1029). If the total effect is larger than the main effect, this input has interaction with others (e.g. P1008 and P1012), the degree of which depends on the difference between main and total effects. The sums of main or total effects for the inputs can also be used to detect interactions. If main effects of all the inputs sum to a value less than but close to 1.0, or if the sum of total effects results in a value greater than but close to 1.0, it can be concluded that the interactions are negligible. Table 3 (last row) indicates that noticeable interactions only exist between P1008 and P1012 for VoidF1 and VoidF2.

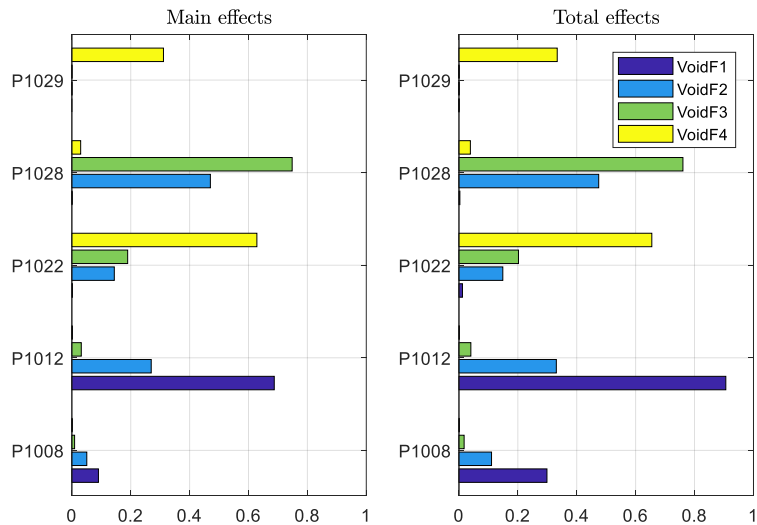

Figure 1: Sobol' indices for selected TRACE physical model parameters, calculated using prior distributions.

Based on the SA results, it can be seen that this numerical test is ideal for investigating the relationship between sensitivity and identifiability. There are multiple inputs and responses, and each response has different significant contributors. For instance, (1) VoidF1 has P1008 and P1012, (2) VoidF2 has P1008, P1012, P1022 and P1028, (3) VoidF3 has P1022 and P1028 (P1012 is negligible), (4) VoidF4 has P1022 and P1029 (P1028 is negligible). During inverse UQ, we can use data from different combinations of the responses and see if the resulting posterior distributions of corresponding significant contributors are identifiable. In this way a connection between sensitivity and identifiability can be established.

**5.2. Posterior distributions and identifiability**

The detailed step-by-step implementation of the improved modular Bayesian approach for this problem can be found in [21]. Therefore, in this paper we will directly present the posterior results pertinent to the identifiability issue. There are 15 different combinations of the four responses, from using only single response to using all of the four responses. Table 4 shows the posterior mean values and STDs for the five physical model parameters and Figure 2 visualizes the results. The "output" column means the combinations of outputs whose data are used for inverse UQ. For instance, output "124" means data from VoidF1, VoidF2 and VoidF4 are used. Taking the SA results into consideration, Table 4 and Figure 2 demonstrate that:

1) P1008 only has small significance for VoidF1 (mostly through interaction with P1012) and even smaller importance for VoidF2. However, it can still be observed that whenever VoidF1 and VoidF2 are included, the posterior STDs of P1008 are relatively smaller. Moreover, including VoidF1 leads to better identifiability than VoidF2 by comparing the results of "1" vs. "2", "13" vs "23", "14" vs "24" and "134" vs "234". For outputs "3", "4" and "34", the posterior STDs of P1008 are much larger than the others.

2) P1012 is significant for VoidF1 and VoidF2. Similar to P1008, the posterior STDs for P1012 are small whenever VoidF1 and VoidF2 are present, and large with outputs "3", "4" and "34". It is also shown that including VoidF1 results in better identifiability than including VoidF2 because P1012 has much higher sensitivity indices for VoidF1.

3) P1022 is significant for VoidF4, and relatively important for VoidF2 and VoidF3. As expected, its posterior STD without any one of these outputs (i.e. outputs "1") is the largest.

4) P1028 is significant for VoidF2 and VoidF3. When either VoidF2 or VoidF3, or both are considered, the posterior STDs are small. Posterior STDs with neither of them (i.e. outputs "1", "4" and "14") are among the largest.

5) P1029 is only significant for VoidF4. Therefore, its posterior STDs are only relatively smaller when VoidF4 is included.

Table 4: Posterior mean values and STDs for TRACE physical model parameters from the first inverse UQ process.

| Output | Posterior mean values | | | | | Posterior STDs | | | | |
|---|---|---|---|---|---|---|---|---|---|---|
| | P1008 | P1012 | P1022 | P1028 | P1029 | P1008 | P1012 | P1022 | P1028 | P1029 |
| 1 | 0.9943 | 1.1422 | 1.7042 | 2.2032 | 1.2974 | 0.5241 | 0.1539 | 0.6474 | 0.5093 | 0.7677 |
| 2 | 1.3339 | 1.1582 | 0.8413 | 1.1614 | 2.7884 | 0.6469 | 0.2378 | 0.4759 | 0.1691 | 1.2545 |
| 3 | 2.7964 | 1.1109 | 1.1060 | 1.1660 | 3.0939 | 1.2038 | 0.6112 | 0.5287 | 0.2944 | 1.2177 |
| 4 | 1.9979 | 2.3113 | 1.6510 | 2.5607 | 1.3588 | 1.2781 | 1.3494 | 0.4418 | 1.2574 | 0.5918 |
| 12 | 0.5854 | 1.2321 | 1.4871 | 1.3385 | 1.7452 | 0.2374 | 0.0921 | 0.3723 | 0.1500 | 0.8096 |
| 13 | 0.5495 | 1.2563 | 1.7204 | 1.4770 | 1.4485 | 0.2414 | 0.1091 | 0.3818 | 0.2698 | 0.8348 |
| 14 | 1.1119 | 1.1215 | 1.5256 | 2.3491 | 1.0182 | 0.3971 | 0.1285 | 0.2604 | 0.4491 | 0.3081 |
| 23 | 1.4224 | 1.1041 | 0.9852 | 1.1117 | 3.5537 | 0.6084 | 0.2510 | 0.3356 | 0.1621 | 0.9919 |
| 24 | 1.0176 | 1.1510 | 1.2568 | 1.3054 | 1.1722 | 0.6200 | 0.2249 | 0.2509 | 0.1476 | 0.3651 |
| 34 | 2.3102 | 0.9528 | 1.4135 | 1.3864 | 1.3342 | 1.1432 | 0.5437 | 0.2477 | 0.2148 | 0.4394 |
| 123 | 0.6201 | 1.2210 | 1.4362 | 1.2921 | 1.8126 | 0.2305 | 0.0920 | 0.2668 | 0.1376 | 0.8596 |
| 124 | 0.5927 | 1.2450 | 1.4012 | 1.3634 | 1.2245 | 0.2088 | 0.0873 | 0.2259 | 0.1498 | 0.3580 |
| 134 | 0.6062 | 1.2497 | 1.5421 | 1.4711 | 1.2252 | 0.2274 | 0.1007 | 0.2412 | 0.2239 | 0.3467 |
| 234 | 1.3945 | 1.0158 | 1.3036 | 1.3258 | 1.2231 | 0.6233 | 0.2099 | 0.2029 | 0.1151 | 0.3735 |
| 1234 | 0.6162 | 1.2358 | 1.4110 | 1.3385 | 1.2340 | 0.2113 | 0.0890 | 0.1833 | 0.1155 | 0.3453 |

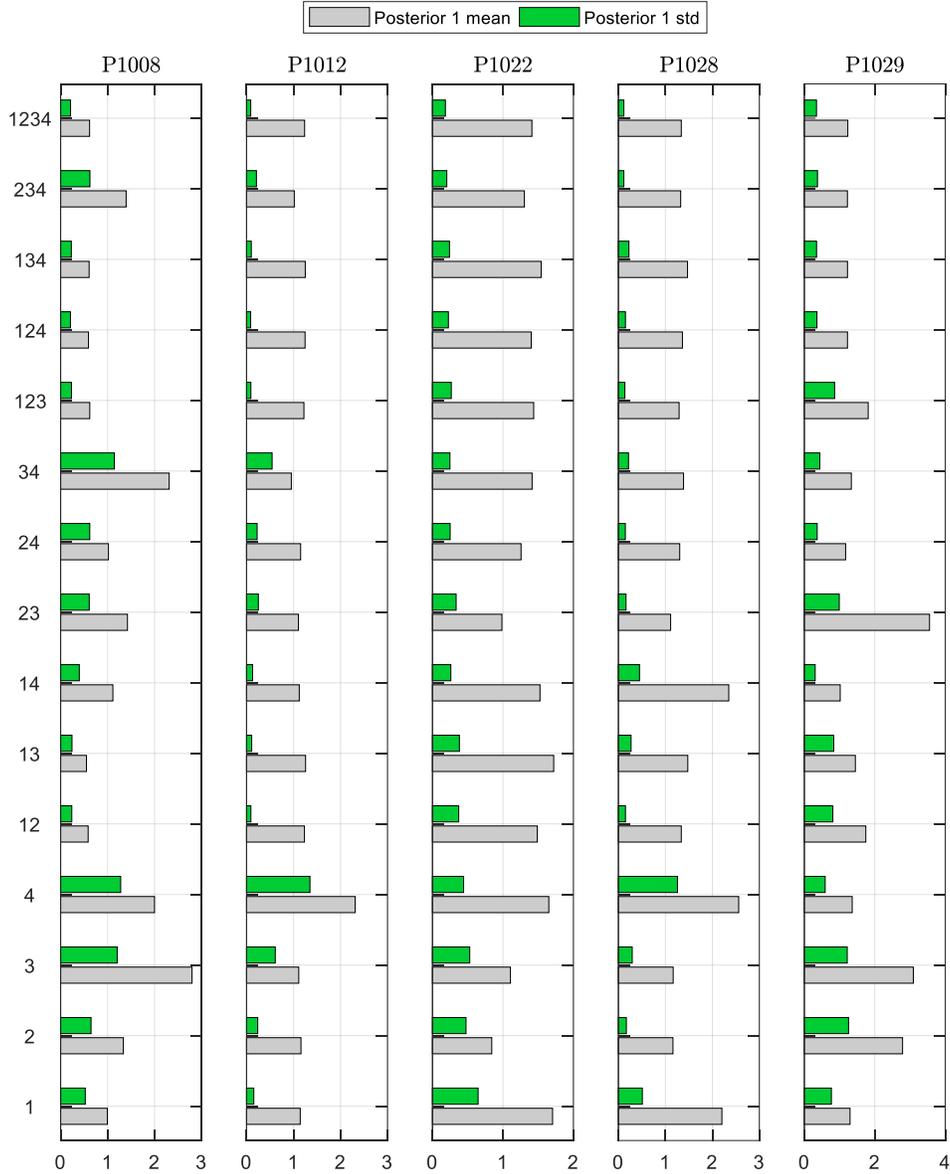

Figure 2: Posterior mean values and STDs for TRACE physical model parameters from the first inverse UQ process.

This numerical test provides solid evidence that identifiability is closely related to sensitivity, which was often ignored in previous research. Good identifiability cannot be achieved for a certain calibration parameter if it is not significant to any of the responses whose data are used for inverse UQ. The posterior distributions obtained with output "1234" are considered as the most appropriate results because all the four responses are considered, and all the five calibration parameters are close to be the most "identifiable" (with nearly smallest posterior STDs) among all the 15 combinations. Figure 3 shows the posterior pair-wise joint densities (off-diagonal sub-figures) and marginal densities (diagonal sub-figures) for the five physical model parameters when all the four responses are used. In Figure 3, the ranges of x-axes of the marginal density plots and the ranges of x/y axes of the joint density plots are the uniform prior ranges. It is obvious that the obtained posterior distributions have substantial reduction in uncertainties compared to the prior distributions. The posterior STDs of P1012 and P1028 are smaller than the other three parameters, indicating larger uncertainty reduction.

The remarkable differences in the posterior statistical moments achieved using different combinations of responses warns us about the risk of obtaining improper posteriors if the observation data is not properly chosen. The first risk is "*bad identifiability*", which is unlikely to be alleviated by numerical techniques (e.g. providing informative priors for model discrepancy). The second risk, however, is "*fake identifiability*", which refers to the case when posterior STDs are small but mean values are wrong. In the introduction part, we have mentioned that misspecified priors can result in tight posterior distributions for calibration parameters which are "far away" from the "true" solutions [6]. Such "fake identifiability" is also possible *if the responses are not properly chosen*. For example, with output "234", the posterior STDs for P1022, P1028 and P1029 are among the smallest possible values, while those for P1008 and P1012 are also acceptably small (44% and 20% of the mean values respectively). However, the mean value for P1008 obtained using output "234" is more than twice of the value obtained using output "1234". Similarly, with output "123", P1008, P1012, P1022 and P1028 are close to be the most identifiable, but the posterior mean value of P1029 is 50% larger than the value obtained with output "1234".

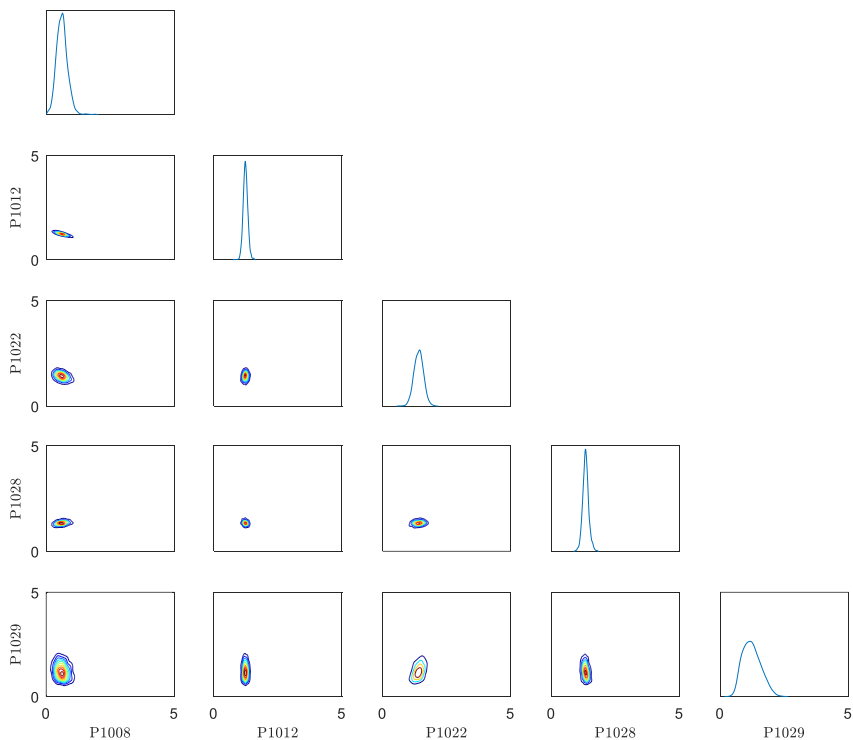

Figure 3: Posterior pair-wise joint and marginal densities when all of the four responses are used.

## 6. Results for the Second Inverse UQ Process

In Section 5.2, identifiability results from a "first" inverse UQ process are shown. Here "first" means that the inverse UQ process starts with the non-informative prior distributions shown in Table 2. In this section, the results for a "second" inverse UQ process will be presented. The motivation for performing another inverse UQ is twofold:

1) Model updating using Bayesian inference is an "iterative" process and the *posterior* distributions achieved from the first inverse UQ process can serve as the *prior* distributions for another inverse UQ process. As shown in Figure 3, the "new prior distributions" are very informative compared to the uniform prior distributions used in the first inverse UQ process. It is worthwhile to perform more inverse UQ "iterations" until the posterior distributions "converge".

2) When the input uncertainties change from non-informative uniform priors (Table 2) to the updated informative priors (Figure 3), Sobol' indices will change accordingly. It is necessary to look at the identifiability of the new posterior results based on the updated Sobol' indices. Substantial uncertainty reduction have been observed for all the parameters. Furthermore, the posterior distributions of P1012 and P1028 have smaller variations than the other three parameters. As a result, their relative importance are expected to reduce.

## 6.1. Sensitivity analysis

Table 5 shows the main and total effects Sobol' indices calculated using the posterior distributions form the first inverse UQ process (Figure 3) and Figure 4 visualizes the results. Compared with Figure 1, it can be seen that:

1) For VoidF1, there are still two significant contributors, P1008 and P1012. However, the relative importance of P1008 and P1012 increases and decreases respectively, due to the larger reduction of uncertainty in P1012. Even though the uncertainty in P1008 is also reduced, its posterior STD is much larger than P1012. High-order interactions still exist between P1008 and P1012 for VoidF1.

2) For VoidF2, P1008 becomes much more important than the others, primarily because of the uncertainty reduction in P1012 and P1028 are larger than P1008, as shown in Figure 3.

3) For VoidF3, the contributions from P1008 and P1022 increase. The main/total effects of P1012 diminish and Sobol' indices of P1012 decrease by nearly half. Similarly, these can explained by the larger reduction of uncertainty in P1012 as shown in Figure 3.

4) Similar results can be observed for VoidF4, there are still two significant contributors, but the importance of P1022 and P1029 have went down/up accordingly. P1029 alone now account for about 60% of the total variation in VoidF4, making it the more significant input.

Table 5: Sobol' indices for TRACE physical model parameters, calculated using posterior distributions from first inverse UQ process.

| Output | Main effect Sobol' indices | | | | Total effect Sobol' indices | | | |
|---|---|---|---|---|---|---|---|---|
| | VoidF1 | VoidF2 | VoidF3 | VoidF4 | VoidF1 | VoidF2 | VoidF3 | VoidF4 |
| P1008 | 0.2581 | 0.6346 | 0.1233 | 0.0041 | 0.5530 | 0.6414 | 0.1230 | 0.0043 |
| P1012 | 0.4757 | 0.0392 | 0.0041 | 0.0002 | 0.7522 | 0.0477 | 0.0044 | 0.0004 |
| P1022 | 0.0055 | 0.1825 | 0.4941 | 0.3835 | 0.0057 | 0.1807 | 0.4919 | 0.3867 |
| P1028 | 0.0006 | 0.1347 | 0.3803 | 0.0104 | 0.0008 | 0.1350 | 0.3809 | 0.0107 |
| P1029 | 0.0000 | 0.0000 | 0.0000 | 0.5991 | 0.0002 | 0.0002 | 0.0002 | 0.6029 |
| Sum | 0.7399 | 0.9910 | 1.0018 | 0.9973 | 1.3119 | 1.0050 | 1.0005 | 1.0050 |

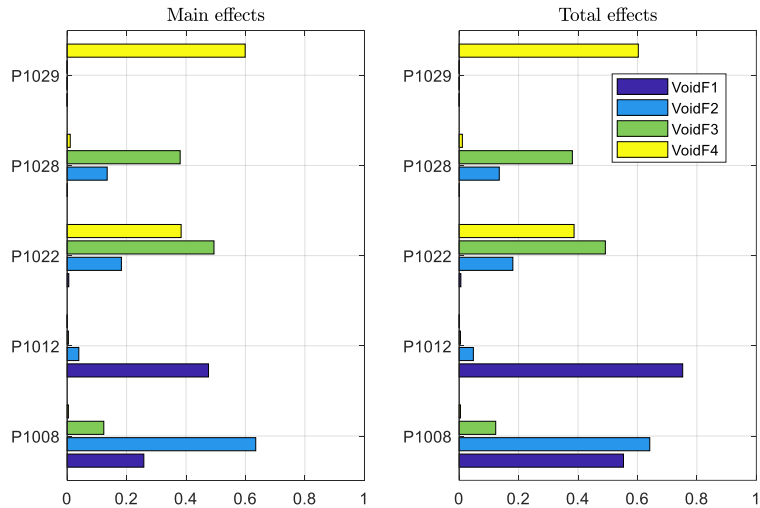

Figure 4: Sobol' indices for physical model parameters, calculated using posterior distributions from first inverse UQ process.

## 6.2. Posterior distributions and identifiability

Table 6 shows the posterior mean values and STDs for the five physical model parameters using different combination of the responses. In Figure 5, the posterior statistical moments from the first (Posterior 1 in the figure) and the second (Posterior 2 in the figure) inverse UQ processes are compared. Note that the Posterior 1 results in Figure 5 are the same with those presented in Figure 2. Here Posterior 2 results are superimposed on Posterior 1 results so that their comparisons are more obvious. The following can be observed from Table 6 and Figure 5:

1) By only looking at the Posterior 2 STDs in Figure 5, there are no notable differences among all the 15 different combinations of responses because they are all very small in magnitude. The reason is that the "updated prior distributions" (which is the posterior distributions from first inverse UQ, as shown in Figure 3) are very informative and they have strong influence on the posterior distributions through Equation (2). As a result, no matter what responses are used, the posterior distributions will not deviate greatly from the "updated prior distributions".

2) However, even though these 15 posterior distributions have similar identifiability by visually inspecting Figure 5, indicated by the nearly equivalent yellow bars, the exact posterior STD values in Table 6 still reflect some connections between sensitivity and identifiability. For instance, P1008 is trivial for VoidF4, its posterior STD is the largest when only VoidF4 is used for inverse UQ (output "4"). P1012 is not important for VoidF3 and VoidF4, and its posterior STDs are large with outputs "3", "4" and "34". P1029 is only significant for P1029, therefore, its posterior STDs are only small when VoidF4 is considered. Similar connections can also be observed for other parameters. It can be concluded that when the prior distributions are extremely informative, the relationship between input significance and posterior identifiability become less obvious but still holds.

3) By comparing the posterior STDs from first (green bars) and second (yellow bars) inverse UQ processes, it can be seen that the posterior distributions from the second inverse UQ process have better identifiability, indicated by their smaller STDs. This study confirms that providing informative priors can indeed improve the identifiability.

Table 6: Posterior mean values and STDs for TRACE physical model parameters from the second inverse UQ process.

| Output | Posterior mean values | | | | | Posterior STDs | | | | |
|---|---|---|---|---|---|---|---|---|---|---|
| | P1008 | P1012 | P1022 | P1028 | P1029 | P1008 | P1012 | P1022 | P1028 | P1029 |
| 1 | 0.5648 | 1.2542 | 1.4289 | 1.3914 | 1.2649 | 0.1278 | 0.0566 | 0.1585 | 0.1039 | 0.3008 |
| 2 | 0.5999 | 1.2451 | 1.3617 | 1.3748 | 1.2734 | 0.1753 | 0.0658 | 0.1535 | 0.0857 | 0.3174 |
| 3 | 0.6514 | 1.2417 | 1.4172 | 1.3687 | 1.2492 | 0.1878 | 0.0812 | 0.1506 | 0.0895 | 0.3290 |
| 4 | 0.6195 | 1.2359 | 1.4607 | 1.3393 | 1.1656 | 0.1966 | 0.0852 | 0.1429 | 0.1052 | 0.2306 |
| 12 | 0.5758 | 1.2543 | 1.3886 | 1.4022 | 1.2953 | 0.1233 | 0.0560 | 0.1508 | 0.0861 | 0.3092 |
| 13 | 0.5782 | 1.2515 | 1.4355 | 1.3990 | 1.2677 | 0.1220 | 0.0554 | 0.1423 | 0.0891 | 0.3062 |
| 14 | 0.5543 | 1.2546 | 1.4706 | 1.3909 | 1.1705 | 0.1214 | 0.0565 | 0.1337 | 0.1044 | 0.2201 |
| 23 | 0.6323 | 1.2412 | 1.3763 | 1.3875 | 1.2846 | 0.1720 | 0.0680 | 0.1458 | 0.0736 | 0.3379 |
| 24 | 0.5831 | 1.2376 | 1.4283 | 1.3798 | 1.1677 | 0.1669 | 0.0657 | 0.1336 | 0.0823 | 0.2379 |
| 34 | 0.6395 | 1.2407 | 1.4603 | 1.3777 | 1.1676 | 0.1851 | 0.0814 | 0.1335 | 0.0851 | 0.2293 |
| 123 | 0.5894 | 1.2505 | 1.4050 | 1.4110 | 1.2939 | 0.1264 | 0.0535 | 0.1411 | 0.0775 | 0.3108 |
| 124 | 0.5569 | 1.2547 | 1.4446 | 1.4120 | 1.2008 | 0.1248 | 0.0551 | 0.1309 | 0.0853 | 0.2253 |
| 134 | 0.5727 | 1.2523 | 1.4806 | 1.4079 | 1.1962 | 0.1195 | 0.0539 | 0.1223 | 0.0905 | 0.2260 |
| 234 | 0.6089 | 1.2381 | 1.4284 | 1.3910 | 1.1790 | 0.1697 | 0.0685 | 0.1250 | 0.0731 | 0.2282 |
| 1234 | 0.5690 | 1.2535 | 1.4480 | 1.4155 | 1.2070 | 0.1260 | 0.0538 | 0.1219 | 0.0768 | 0.2382 |

Figure 6 compares the posterior marginal and joint distributions from first (solid lines) and second (dash-dotted lines) inverse UQ processes. The pair-wise joint distributions of both posteriors show similar dependence structure, while the second posterior joint distributions are more concentrated. By comparing the marginal distributions, both posterior distributions have similar shapes, while the second posterior marginal distributions have smaller variances. These facts again indicate that the posterior distributions from the second inverse UQ process are more "identifiable", which is due to the much more informative "prior" distributions. Note that performing more inverse UQ "iterations" does make the posterior distributions more concentrated. Therefore, in this paper posterior results from further inverse UQ process will not be reported.

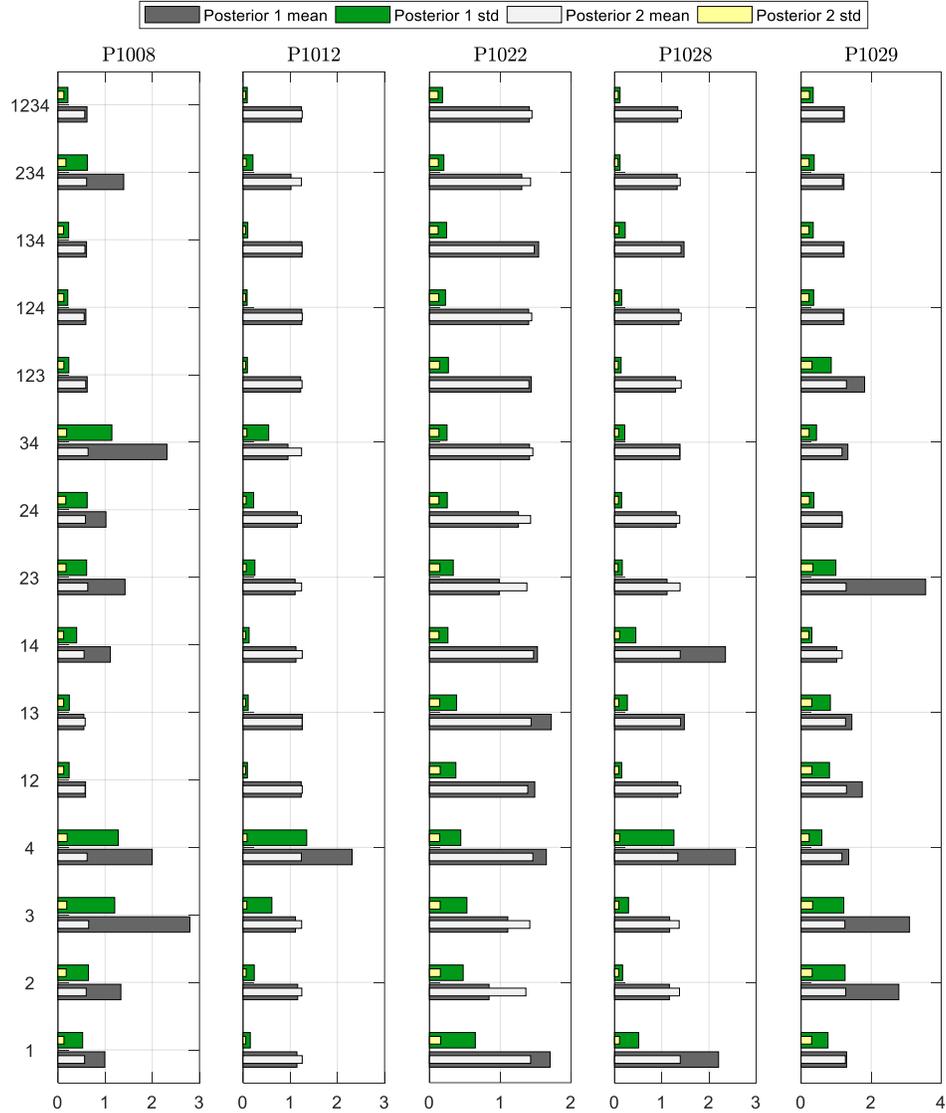

Figure 5: Comparison of posterior mean values and STDs for TRACE physical model parameters from first (Posterior 1) and second (Posterior 2) inverse UQ process.

This numerical example also demonstrates the danger of getting "fake identifiability" in case of misspecified informative priors, because the resulting posterior will tend to "converge" to the misspecified informative priors. In the second inverse UQ process, those informative priors are posterior distributions obtained from the first inverse UQ process. Imagine that some *inaccurate but informative priors* are used which can deviate greatly from the "true" values of the calibration parameters. Similar to the second posterior in Figure 6, the corresponding posterior distributions will also "converge" to the inaccurate but informative prior distributions and their STDs will be small. In this case, one may believe that good identifiability has been achieved when the posterior distributions are hardly "close" to the "true" values.

Through the numerical study results in Section 5.2 and 6.2, it has been demonstrated that fake identifiability is possible if model responses are not appropriately chosen, or inaccurate but informative priors are specified. Therefore, we recommend that during real applications of inverse UQ, one should first conduct a sensitivity study to identify the important calibration parameters for the responses whose data are available. Next non-informative prior distributions should be selected for those non-trivial calibration parameters. One can then perform a few Inverse UQ "iterations" until the posterior distributions "converge".

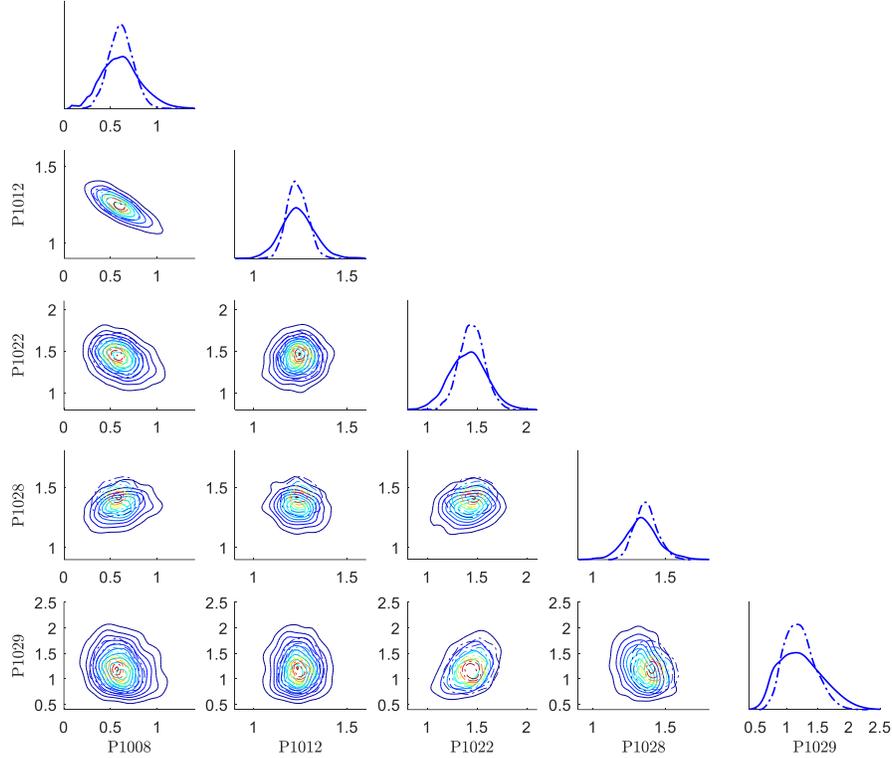

Figure 6: Comparison of posterior pair-wise joint and marginal densities from first (solid lines) and second (dash-doted lines) inverse UQ process, when all of them four responses are used.

**7. Conclusions**

In this work, the main objective is to demonstrate the relationship between sensitivity and identifiability for inverse UQ. The identifiability issue is caused by the confounding between the calibration parameters and the model discrepancy. Several different combinations of the computer model (with different values for the calibration parameters) and its corresponding model discrepancy might result in equally good agreement with the measurement data and equally high values for the likelihood function during MCMC sampling. Consequently, the calibration parameters become non-identifiable. Previous work have been focusing on providing informative priors for the calibration parameters and/or the model discrepancy function. However, this is a crude and often impossible solution to improve identifiability because one rarely has such informative prior knowledge.

We adopted an improved modular Bayesian approach for inverse UQ that does not require priors for the model discrepancy term. We investigated a practical example in nuclear engineering, which is the inverse UQ of nuclear reactor system thermal-hydraulics code TRACE physical model parameters using BFBT benchmark data. This test consists of five calibration parameters and four responses with different significant contributors. By performing inverse UQ with different combinations of the responses and looking at the corresponding posteriors, a connection between sensitivity and identifiability has been established.

It has been shown that identifiability is largely related to the sensitivity (or significance) of the calibration parameters w.r.t. the chosen responses. The numerical test demonstrated that, in order for a certain calibration parameter to be statistically identifiable, it should be significant to at least one of the responses whose data are used for inverse UQ. Good identifiability cannot be achieved for a certain calibration parameter if it is not significant to any response. Therefore, to obtain good identifiability in inverse UQ or Bayesian calibration activities, we recommend that the users first conduct a sensitivity study to identify the important calibration parameters for the responses whose data are available. Next non-informative prior distributions should be selected for those non-trivial calibration parameters. One can then perform a few Inverse UQ "iterations" until the posterior distributions "converge". In this way, the risk of both bad identifiability and fake identifiability can be reduced.